\documentclass[letterpaper,titlepage,11pt]{article}
\usepackage{hyperref}
\usepackage{amssymb,amsmath,amsfonts}
\usepackage{epsfig}
\def\clock{{\count0=\time
           \divide\count0 60
           \ifnum\count0<10 0\fi\the\count0
           \multiply\count0 -60 \advance\count0 \time
           :\ifnum\count0<10 0\fi \the\count0
         }}
\newcommand{\timestamp}{{\small\vbox{\hbox{\tt\jobname.tex}
\hbox{\the\day/\the\month/\the\year, \clock}}}}

\newcommand{\CF}{\mathcal{F}}

\newcommand{\CO}{\mathcal{O}}

\newcommand{\spa}{\ , \ \ }

\newcommand{\ds}{\displaystyle}

\numberwithin{equation}{section}

\begin{document}

\begin{titlepage}

\rightline{\vbox{\small\hbox{\tt NORDITA-2011-3} }}
 \vskip 1.4 cm

\centerline{\Huge \bf Thermodynamics of the hot BIon}
\vskip 1.5cm

\centerline{\large {\bf Gianluca Grignani$\,^{1}$},  {\bf Troels Harmark$\,^{2}$},}
\vskip 0.2cm \centerline{\large  {\bf Andrea Marini$\,^{1}$} ,  {\bf Niels A. Obers$\,^{3}$} and
{\bf Marta Orselli$\,^{3}$} }

\vskip 0.8cm

\begin{center}
\sl $^1$ Dipartimento di Fisica, Universit\`a di Perugia,\\
I.N.F.N. Sezione di Perugia,\\
Via Pascoli, I-06123 Perugia, Italy
\vskip 0.2cm
\sl $^2$ NORDITA\\
Roslagstullsbacken 23,
SE-106 91 Stockholm,
Sweden \vskip 0.2cm
\sl $^3$ The Niels Bohr Institute  \\
\sl  Blegdamsvej 17, DK-2100 Copenhagen \O , Denmark
\end{center}
\vskip 0.4cm

\centerline{\small\tt grignani@pg.infn.it, harmark@nordita.org, }
\centerline{\small\tt andrea.marini@fisica.unipg.it, obers@nbi.dk, orselli@nbi.dk}

\vskip 1.1cm \centerline{\bf Abstract} \vskip 0.2cm \noindent
We investigate the thermodynamics of the recently obtained finite temperature BIon solution of arXiv:1012.1494, focusing on two aspects.
The first concerns comparison of the free energy of the three available
phases for the finite temperature brane-antibrane wormhole configuration.
Based on this we propose a heuristic picture for the dynamics of the phases
that involves a critical temperature below which a stable phase exists.
This stable phase is the finite temperature analogue of the thin throat branch
of the extremal brane anti-brane wormhole configuration.
The second aspect that we consider is the possibility of constructing a finite temperature generalization of the infinite spike configuration of the extremal BIon. To this end  we identify a
correspondence point at the end of the throat where the thermodynamics
of the D3-F1 blackfold configuration can be matched to that of $k$ non-extremal black fundamental strings.

\end{titlepage}

\small
\tableofcontents
\normalsize
\setcounter{page}{1}

\section{Introduction}

In a recent paper~\cite{Grignani:2010xm} we proposed a novel method to describe D-branes probing thermal backgrounds. This was based on the blackfold approach developed in~\cite{Emparan:2007wm}. The main feature of this new method is that it describes a brane probe consisting of a large number of coincident non-extremal D-branes such that the probe is in thermal equilibrium with the background. This new method was employed in \cite{Grignani:2010xm} to heat up the BIon solution~\cite{Callan:1997kz,Gibbons:1997xz} by putting it in the background of hot flat space. Doing this, it was found that the finite temperature BIon behaves qualitatively different than its zero-temperature counterpart.

A key observation of~\cite{Grignani:2010xm} is that the requirement of thermal equilibrium between the probe and the background changes the probe not only globally but also locally in that it changes the energy-momentum tensor in the equations of motion of the probe. This is due to the fact that the degrees of freedom living on the brane gets heated up. While this is taken into account in our new method it is, however, not taken into account in the previously used approach for describing D-brane probes in thermal background which employed the classical Dirac-Born-Infeld (DBI) action. Thus, the previously used approach does not provide an accurate description of D-brane probes in thermal backgrounds.%
\footnote{See~Section 3.3 of \cite{Grignani:2010xm} for a more elaborate and precise argument for this.}
Our method could thus potentially open up new insights and developments in the study of D-branes as probes of thermal backgrounds particularly with applications to the AdS/CFT correspondence.

In this paper we continue our study of the finite temperature BIon solution found in \cite{Grignani:2010xm} which generalizes the extremal BIon
solution in the particular case of the D3-brane. The extremal BIon solution either takes the form of an infinite spike with an F-string charge coming out of the D3-brane, describing a number of coincident F-strings ending on the D3-brane, or the form of a D3-brane and a parallel anti-D3-brane connected by a ``wormhole" with F-string charge which we dub the brane-antibrane-wormhole configuration. The latter solution would correspond in the linearized regime to a string stretched between a D3-brane and an anti-D3-brane. The finite temperature BIon solution of \cite{Grignani:2010xm} is a generalization of the brane-antibrane-wormhole configuration which is found employing a non-extremal D3-F1 probe brane system curved in hot flat space. Note that, while the extremal BIon solution of the DBI action is found for a system with one D3-brane (and anti-D3-brane) with $g_s \ll 1$ the finite temperature BIon solution is instead found in a regime with $N$ D3-branes (and anti-D3-branes) with $N \gg 1$ and $g_s N \gg 1$.

We focus in this paper on two aspects of the finite temperature BIon solution which we did not address in \cite{Grignani:2010xm}. The first concerns the structure of the phases of the brane-antibrane-wormhole configuration at finite temperature. It was found in \cite{Grignani:2010xm} that there are up to three available phases for a given temperature and separation between the D3-branes and anti-D3-branes, in contrast with the two available phases for the extremal BIon. Here we examine this further by comparing the free energy in the appropriate thermodynamical ensemble to see which available phase is the dominant one. Moreover, based on this, we propose a heuristic picture for the dynamics of the phases. The second aspect that we consider in this paper is the possibility of constructing a finite temperature generalization of the infinite spike configuraton of the extremal BIon.

\begin{figure}[h!]
\centerline{\includegraphics[scale=0.4]{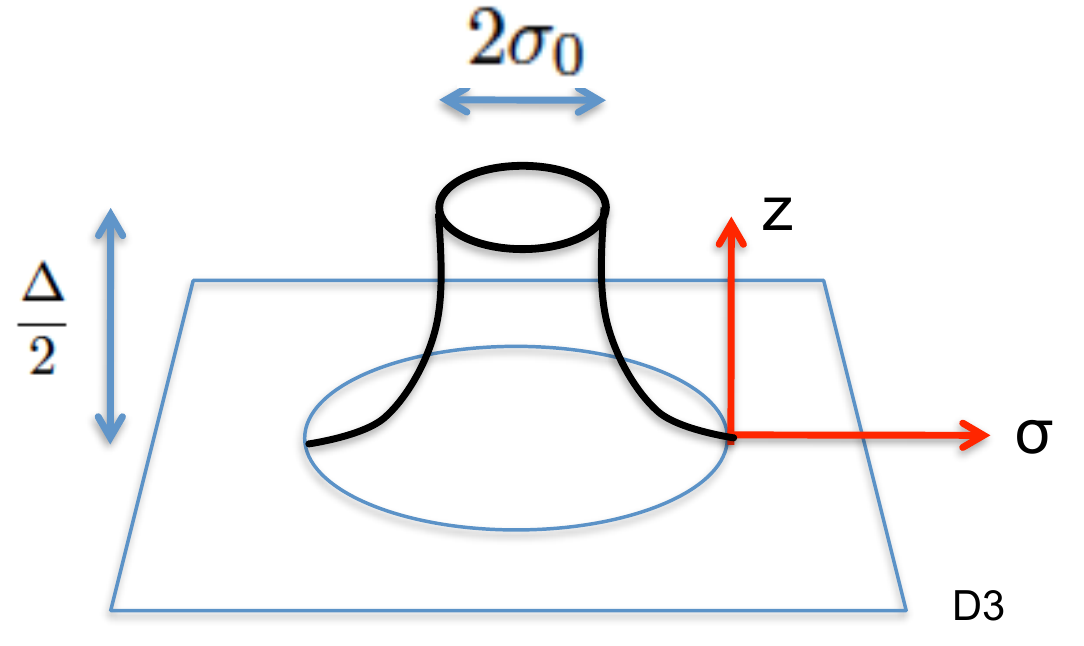}}
\caption{\small Illustration of the setup, showing the embedding function $z(\sigma)$ and
the definition of the parameters $\sigma_0$ and $\Delta$.}
\label{fig:setup}
\end{figure}

We begin by reviewing here the main features of the finite temperature BIon solution \cite{Grignani:2010xm}. The geometric setup is as follows. Start with $N$ coincident flat non-extremal D3-branes embedded in flat space with spherical symmetry in the world-volume directions. Let $z$ be a transverse coordinate to the branes and let $\sigma$ be the radius on the world-volume. The case of the flat branes is thus $z(\sigma)=0$. The curving of the $N$ D3-branes is then described by the profile $z(\sigma)$, as illustrated in Fig.~\ref{fig:setup}. We impose the two boundary conditions that $z(\sigma)\rightarrow 0$ for $\sigma\rightarrow \infty$ and $z'(\sigma) \rightarrow - \infty$ for $\sigma \rightarrow \sigma_0$, where $\sigma_0$ is the minimal two-sphere radius of the configuration. Putting now $k$ units of F-string charge along the radial direction we obtain the profile
\begin{equation}
\label{thesolution}
z(\sigma) = \int_{\sigma}^\infty d{\sigma'} \left( \frac{F({\sigma'})^2}{F(\sigma_0)^2} - 1 \right)^{-\frac{1}{2}}
\end{equation}
where $F(\sigma)$ is given by
\begin{equation}
\label{Ffunction}
F(\sigma) = \sigma^2  \frac{ 4 \cosh^2 \alpha -3 }{\cosh^4 \alpha}
\end{equation}
In \cite{Grignani:2010xm} we found that there are two possible branches of solutions determining the function $\alpha(\sigma)$, one being the branch connected to the extremal BIon, the other being connected to a neutral 3-brane with F-string charge. In this paper we only consider the former branch, in which case we have
\begin{equation}
\label{mainbranch}
\cosh^2 \alpha = \frac{3}{2} \frac{\cos \frac{\delta}{3} + \sqrt{3} \sin \frac{\delta}{3} } {\cos \delta}
\end{equation}
with the definitions%
\footnote{Here the F-string tensions is $T_{\rm F1} = 1 / (2\pi l_s^2 )$ and the D3-brane tension is $T_{\rm D3} = 1/((2\pi)^3 g_s l_s^4 )$.}
\begin{equation}
\label{T_kappa_delta}
\cos \delta (\sigma) \equiv \bar{T}^4 \sqrt{1+ \frac{\kappa^2}{\sigma^4}}
\spa
\bar{T} \equiv  \left( \frac{ 9 \pi^2 N}{4\sqrt{3} T_{\rm D3}} \right)^{\frac{1}{4}}  T
\spa
\kappa \equiv \frac{k T_{\rm F1}}{4\pi N T_{\rm D3}}
\end{equation}
We see from the first equation above that the minimal radius $\sigma_0$ is bounded from below $\sigma_0 \geq \sigma_{\rm min} \equiv \sqrt{\kappa} \bar{T}^2 (1- \bar{T}^8)^{-1/4}$. Note also that $0\leq \bar{T} \leq 1$. From the above solution \eqref{thesolution} we can now construct the brane-antibrane-wormhole configuration by attaching a mirror of the solution, reflected in the plane $z=z(\sigma_0)$. The separation distance $\Delta$ between the $N$ D3-branes and the $N$ anti-D3-branes is then given by $\Delta = 2z(\sigma_0)$ as measured far away from the wormhole, $i.e.$ for $\sigma \gg \sigma_0$ (see Fig.~\ref{fig:setup}).

\begin{figure}[h!]
\centerline{\includegraphics[scale=0.8]{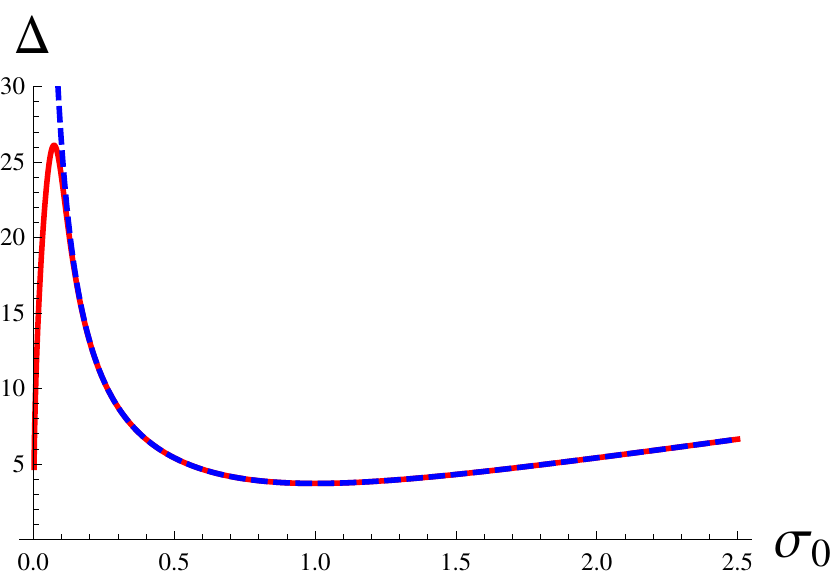} \includegraphics[scale=0.8]{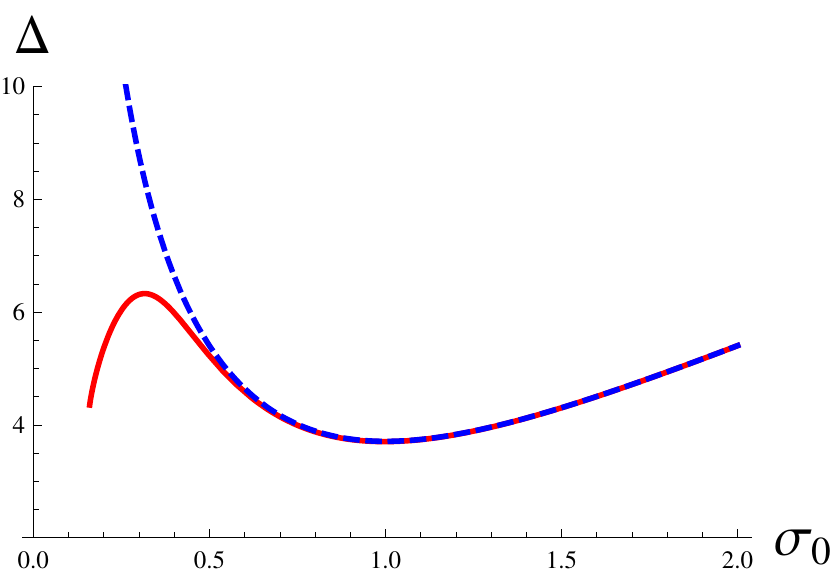}}
\centerline{\includegraphics[scale=0.8]{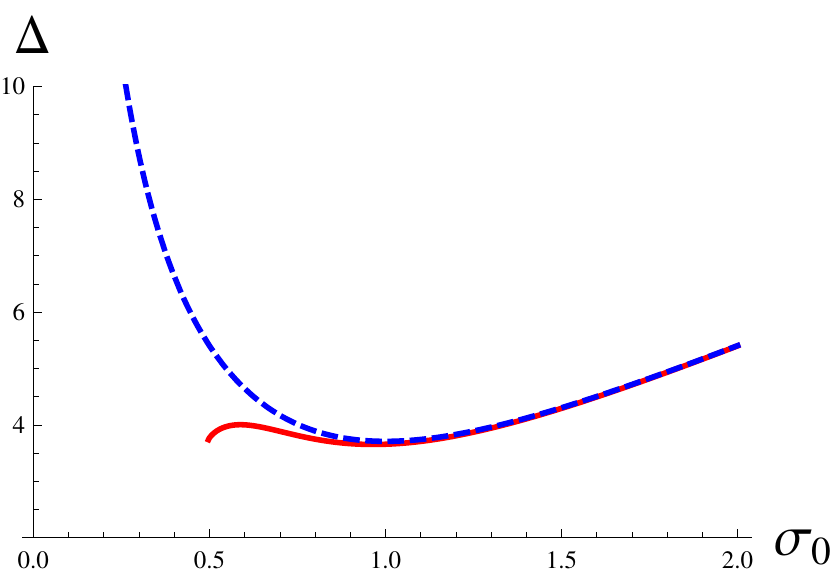} \includegraphics[scale=0.8]{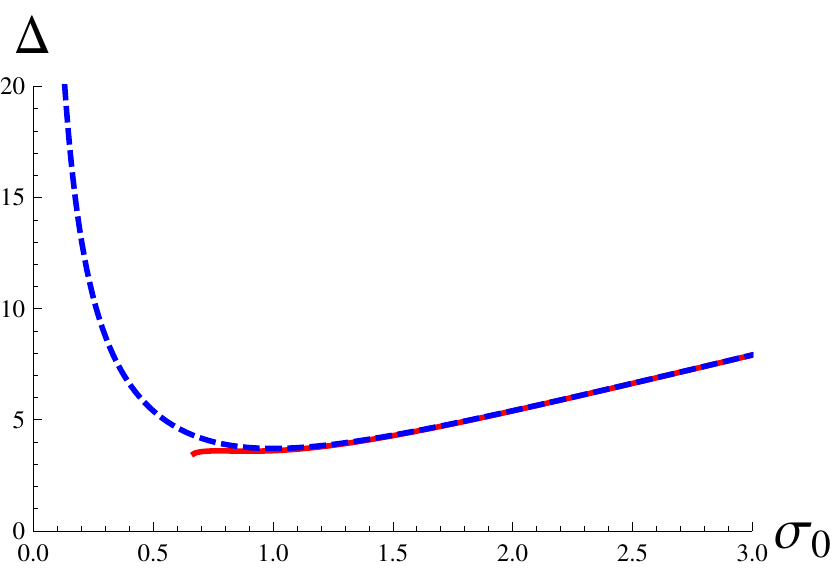}}
\caption{\small On the figures the solid red line is $\Delta$ versus $\sigma_0$ for  $\bar{T}=0.05$ (top left figure), $\bar{T}=0.4$ (top right figure), $\bar{T}=0.7$ (bottom left figure) and $\bar{T}=0.8$ (bottom right figure) while the blue dashed line corresponds to $\bar{T}=0$. We have set $\kappa=1$.}
\label{DeltaPlots}
\end{figure}

In \cite{Grignani:2010xm} we analyzed the separation distance $\Delta$ as a function of the minimal radius $\sigma_0$ for a given temperature $\bar{T}$. The results of this analysis are most easily illustrated by the Figs.~\ref{DeltaPlots} where we plotted $\Delta$ versus $\sigma_0$ for $\bar{T} = 0.05, 0.4, 0.7, 0.8$. Note that we set $\kappa=1$ without loss of generality since we can reinstate a general $\kappa$ by the transformation $\sigma_0 \rightarrow \sqrt{\kappa} \sigma_0$ and $\Delta \rightarrow \sqrt{\kappa} \Delta$. For comparison the figures also include the $\Delta$ versus $\sigma_0$ curve for the zero temperature case, for which the wormhole solution is characterized by a ``thin throat" branch with small $\sigma_0$ and a ``thick throat" branch with large $\sigma_0$ for fixed $\Delta$. Instead when the temperature is turned on, the separation distance $\Delta$ between the brane-antibrane system develops a local maximum in the region corresponding to the zero temperature thin branch. This is a new feature compared to the zero temperature case. The existence of this maximum gives rise to three possible phases with different $\sigma_0$ for a given $\Delta$. For small temperatures and/or large $\sigma_0$ the $\Delta$ as a function of $\sigma_0$ resembles increasingly closely the zero temperature counterpart.

\begin{figure}[ht]
\centering
\includegraphics[scale=0.9]{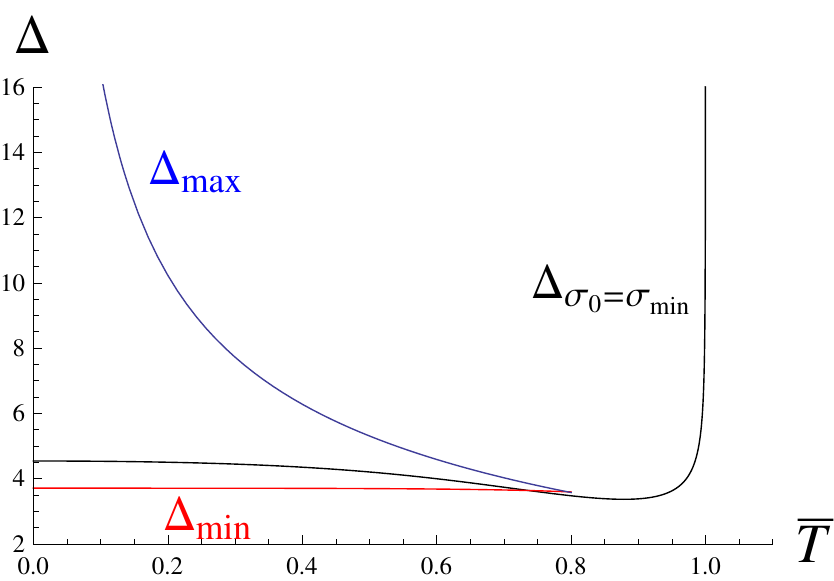}
\caption{$\Delta_{\rm max}$ (blue curve), $\Delta_{\rm min}$ (red curve), $\Delta$
 at $\sigma_{\rm min}$ (black curve) as a function of the temperature $\bar T$ for $\kappa=1$.}
\label{deltaT}
\end{figure}

In the following we denote the value of $\Delta$ at the local maximum as $\Delta_{\rm max}$ and $\Delta$ at the (local) minimum as $\Delta_{\rm min}$. Studying the $\Delta$ versus $\sigma_0$ for small temperatures we found analytically that $\Delta_{\rm max} \propto \sqrt{\kappa} \bar{T}^{-2/3}$ while $\Delta_{\rm min} \propto \sqrt{\kappa} \bar{T}^0$. In Fig.~\ref{deltaT} we plotted $\Delta_{\rm max}$ and $\Delta_{\rm min}$, along with $\Delta$ at $\sigma_0=\sigma_{\rm min}$, for the whole range of temperatures $0\leq \bar{T} \leq 1$. We see from this that there is a critical value of $\bar{T}$ given by $\bar{T}_b \simeq 0.8$ beyond which $\Delta_{\rm max}$ and $\Delta_{\rm min}$ cease to exist.

Given the three phases described above, a natural question to ask is which of the phases
will dominate. This aspect is studied in Sec.~\ref{sec:freeenergy} of the paper, where we consider
the behavior of the finite-temperature brane-antibrane wormhole system in the canonical ensemble.
We  first define in Sec.~\ref{sec:deffreeenergy} the free energy relative to a cutoff value of $\sigma_0$ and show
that there is no dependence on the cutoff when measuring the relative difference in the free energy.
When comparing the free energies, the resulting picture can be summarized as follows.
For a fixed temperature $T$ (below $T_{\rm b}$)
we find that for brane separations below $\Delta_{\rm max}$ the phase with $d \Delta/d \sigma_0 <0$
has lowest free energy and is the dominating branch. This may be considered as the finite temperature
analogue of the thin throat branch in the extremal case. We refer to this as the finite temperature thin throat branch below.
For brane separation larger than $\Delta_{\rm max}$
there is only one available branch (the finite temperature thick throat branch) and we argue that this corresponds to an unstable saddle point.
An alternative way to look at the system is to fix the separation distance and vary the temperature.
From this viewpoint we find that there is a critical temperature $T_{\rm c}$ below which
the finite temperature thin throat branch is the dominating phase. Above the critical temperature
only the finite temperature thick throat branch is available. For comparison we consider in Sec.~\ref{sec:comphas} the
energy in the extremal case, which shows that in this case the thin throat branch is always the energetically
favored configuration. Finally, we present in Sec.~\ref{sec:extren} a heuristic picture
of the system away from equilibrium that forms the basis for our argument that the finite temperature
thick throat branch is an unstable saddle point.

The thermal throat solution found in \cite{Grignani:2010xm} does not allow for an infinite spike. One may thus wonder whether it is possible to construct a thermal spike, i.e. a configuration of $k$ coincident F-strings ending on $N$ coincident D3-branes at non-zero temperature.%
\footnote{This could be relevant for applying our method to \cite{Hartnoll:2006hr}.}
At zero temperature this configuration for $N = 1$ is described by the infinite spike BIon of \cite{Callan:1997kz}.
This question is addressed in Sec.~\ref{sec:spike} of this paper.  We propose that the configuration in question can be constructed by matching up a non-extremal black string solution with the throat solution  \eqref{thesolution}-\eqref{T_kappa_delta}.  In this description, the end of the
throat can thus be viewed as a correspondence point where the thermodynamics
of the D3-F1 blackfold configuration can be matched to that of $k$ non-extremal black F-strings. This matching requires an extrapolation of
the two descriptions into a regime that goes beyond the validity of the descriptions. We argue, however,
that while this poses a limitation to the approach in this case, at the same time it may be regarded as a clear indication that the match between the D3-F1 blackfold configuration and the non-extremal F-strings is possible.

Both of the two aspects that we examine in this paper rely on the detailed properties of the
thermodynamic quantities of the finite temperature throat solution \eqref{thesolution}-\eqref{T_kappa_delta}.
These were obtained in Ref.~\cite{Grignani:2010xm} and are given by the following expressions in terms of the three parameters $(T,N,k)$. For the brane-antibrane wormhole configuration
the total mass, entropy and free energy are
\begin{equation}
\label{Msolution}
M =\frac{4 T_{\rm D3}^2}{\pi T^4} \int_{\sigma_0}^\infty d \sigma
        \frac{F(\sigma)\sigma^2}{ \sqrt{F^2(\sigma)-F^2(\sigma_0)} }  \frac{4 \cosh^2 \alpha+ 1}{\cosh^4 \alpha} \spa
\end{equation}
\begin{equation}
\label{Ssolution}
S =   \frac{4 T_{\rm D3}^2}{\pi T^5} \int_{\sigma_0}^\infty d \sigma
        \frac{F(\sigma)\sigma^2}{\sqrt{F^2(\sigma)-F^2(\sigma_0)} }  \frac{4}{\cosh^4 \alpha}
\end{equation}

\begin{equation}
\label{Fsolution}
\CF = \frac{4 T_{\rm D3}^2}{\pi T^4}   \int_{\sigma_0}^\infty d\sigma \sqrt{1+z'(\sigma)^2}
F(\sigma)
\end{equation}
The D3-brane and F-string chemical potential are given by
\begin{equation}
\label{mu3solution}
\mu_{\rm D3} = 8 \pi T_{\rm D3}  \int_{\sigma_0}^\infty d \sigma
        \frac{F(\sigma)}{\sqrt{F^2(\sigma)-F^2(\sigma_0)} } \,\sigma^2
         \tanh \alpha \cos \zeta
\end{equation}
\begin{equation}
\label{mu1solution}
\mu_{\rm F1} = 2T_{\rm F1} \int_{\sigma_0}^\infty d \sigma
        \frac{F(\sigma)}{\sqrt{F^2(\sigma)-F^2(\sigma_0)} } \,
         \tanh \alpha \sin \zeta
\end{equation}
These quantities satisfy the first law of thermodynamics and Smarr relation
\begin{equation}
\label{firstlaw}
d M = T d S + \mu_{\rm D3} d N +  \mu_{\rm F1} d k \spa
4 (M- \mu_{\rm D3} N - \mu_{\rm F1} k) = 5 TS
\end{equation}
which can be checked explicitly from the expressions above.
Note that for the throat solution ($i.e.$ without attaching a mirror solution) all integrated quantities
should be divided by two.

We conclude this introduction by pointing out that while this paper treats a specific case, it should be emphasized that our new method has a very wide range of applicability. Indeed, the results of this paper and of Ref.~\cite{Grignani:2010xm} show that our method for describing D-branes probing thermal backgrounds is able to reveal new insights into physics of finite-temperature string theory.

\section{Comparison of phases in canonical ensemble}
\label{sec:freeenergy}

In this section we study the physics of the three different phases in the brane-antibrane wormhole configuration found in~\cite{Grignani:2010xm}. We define the free energy of the system and use this to compare the three phases in order to see which is the thermodynamically preferred one. We consider what happens when changing the temperature or the separation distance in the system. Furthermore, we compare the free energy to the energy of the phases in the zero-temperature case of the BIon solution~\cite{Callan:1997kz,Gibbons:1997xz}, see brief review in Section 2 of \cite{Grignani:2010xm}. Finally, we provide some heuristic considerations about the dynamics of the system away from equilibrium.

\subsection{Choice of ensemble and measurement of the free energy}
\label{sec:deffreeenergy}

As reviewed in the Introduction, in~\cite{Grignani:2010xm} we found that the $(\sigma_0,\Delta)$ diagram has important qualitative differences with the $(\sigma_0,\Delta)$ diagram of the BIon solution of the Born-Infeld theory. Notably, for temperatures not too large there exists three distinct phases, $i.e.$ three possible values of $\sigma_0$, for a range of separation distances $\Delta$. It is therefore a natural question which of the three phases are preferred thermodynamically. Before venturing into such a comparison, we should briefly ponder on which thermodynamic ensemble one should consider, $i.e.$ which quantities to keep fixed in the comparison.
We imagine here that the two systems of $N$ D3-branes and $N$ anti-D3-branes are infinitely extended, at least in comparison with the range of values of $\sigma_0$ that we will consider. Two obvious quantities to keep fixed are thus the D3-brane charge (which can be measured at each point of the D3-branes) and the temperature $T$. This is simply because it would take an infinite amount of energy to change the D3-brane charge or the temperature.%
\footnote{One can of course change the temperature locally but we are here concerned with systems in global thermodynamical equilibrium.}
With respect to the temperature we can say that the infinite extension of the branes means that the branes, up to some distance away from the wormhole, correspond to a heat bath for the part of the brane system with the wormhole with $k$ units of F-string flux that connects the two infinitely extended branes. Therefore we are obviously in the canonical ensemble. In addition, also the number of F-strings $k$ should be fixed. This is because if we consider a sufficiently large value of $\sigma$ so that we are far away from the part with the wormhole then we can still measure the $k$ F-strings by making an integral over the two-sphere with radius $\sigma$. From the charge conservation along the F-string world-volume direction we then get that $k$ should be kept fixed in our thermodynamical ensemble. Finally, it is also natural to keep fixed $\Delta$, the separation between the D3-branes and anti-D3-branes, since that is also something we can measure far away from the part of the system where the branes are connected with a wormhole.

In summary, we work in an ensemble with $T$, $N$, $k$ and $\Delta$ kept fixed. The first law of thermodynamics \eqref{firstlaw} generalizes to $dM = T dS + \mu_{\rm D3} dN + \mu_{\rm F1} dk + f d\Delta$ when we allow for variations of $\Delta$ where $f$ is the force between the D3-brane systems. We see from this that the correct free energy for our ensemble is $\CF = M - TS$ with the variation $d\CF = - S dT + \mu_{\rm D3} dN + \mu_{\rm F1} dk + f d\Delta$. Thus, we can write $\CF( T, N, k , \Delta)$. However, as shown in \cite{Grignani:2010xm}, we can have up to three distinct phases given $( T, N, k , \Delta)$ which we can label by $\sigma_0$. Thus, we write below $\CF( T, N, k , \Delta ; \sigma_0)$ where $\sigma_0$ is understood only as a label that points to which branch we are on, and should not be understood as a thermodynamic variable. From \eqref{thesolution}-\eqref{T_kappa_delta} and \eqref{Fsolution} we find
\begin{equation}
\label{freeenergy}
\CF(T,N,k,\Delta ; \sigma_0) = \frac{4 T_{\rm D3}^2}{\pi T^4}   \int_{\sigma_0}^\infty d\sigma  \frac{F(\sigma)^2}{\sqrt{F(\sigma)^2 - F(\sigma_0)^2}}
\end{equation}
The goal is now to compute \eqref{freeenergy} for the various phases for different temperatures and brane separations.

In our application we shall keep $N$ and $k$ strictly fixed. Thus, as such, there is no reason to consider them as variables.
Note that, from Eqs.~\eqref{Ffunction}, \eqref{mainbranch} and \eqref{T_kappa_delta}, $F(\sigma)$ only depends on $N$, $k$ and $T$ through the variables $\kappa$ and $\bar{T}$.
Furthermore, changing $\kappa$ only amounts to a uniform rescaling of the system. Indeed, it is easy to show that finding a solution with a given $\bar{T}$, $\sigma_0$, $\Delta$ and $\CF$ for $\kappa=1$ the general $\kappa$ configuration is found by rescaling $\sigma_0 \rightarrow \sqrt{\kappa} \sigma_0$, $\Delta \rightarrow \sqrt{\kappa} \Delta$ and $\CF \rightarrow \kappa^{3/2} \CF$ while keeping $\bar{T}$ fixed. Thus, we choose to set $\kappa=1$ in the rest of this section since one can always reinstate it by a trivial rescaling and since we are not interested in varying $\kappa$. In accordance with this, we shall consider the variation of $\CF$ with respect to the rescaled temperature $\bar{T}$ rather than $T$. With this we can write \eqref{freeenergy} as
\begin{equation}
\label{freeenergy2}
\CF(\bar{T},\Delta ; \hat{\sigma}_0) = \frac{9\pi N T_{\rm D3}}{\sqrt{3}  \bar{T}^{4} }  \int_{{\sigma}_0}^\infty d{\sigma}  \frac{F(\bar{T},{\sigma})^2}{\sqrt{F(\bar{T},{\sigma})^2 - F(\bar{T},{\sigma_0})^2}}
\end{equation}
where $F(\bar{T},\sigma)$ is defined as the function $F(\sigma)$ given by \eqref{Ffunction}, \eqref{mainbranch} and \eqref{T_kappa_delta} for $\kappa=1$ and a given $\bar{T}$. For our purposes below we furthermore choose to set $N T_{\rm D3}$ to one, $i.e.$ we will measure the free energy in units of $N T_{\rm D3}$. We can thus write
\begin{equation}
\label{freeenergy3}
\CF(\bar{T},\Delta ; {\sigma}_0) = \int_{{\sigma}_0}^\infty d{\sigma}\,  h(\bar{T},{\sigma},{\sigma}_0) \spa h(\bar{T},{\sigma},{\sigma}_0) \equiv    \frac{9\pi F(\bar{T},{\sigma})^2}{\sqrt{3} \bar{T}^{4}\sqrt{F(\bar{T},{\sigma})^2 - F(\bar{T},{\sigma}_0)^2}}
\end{equation}
Consider now the integrand $h(\bar{T},{\sigma},{\sigma}_0)$ in \eqref{freeenergy3}.
Using that in the regime of large $\sigma/\sqrt{\kappa}$ the function $F(\sigma)$ behaves as~\cite{Grignani:2010xm}
\begin{equation}
\label{Flargesigma}
F(\sigma)
=\sigma^2g(\bar{T})+\CO(\kappa/\sigma^2)
\end{equation}
where $g(\bar{T})$ is a function that increases from 0 to $4/3$ as $\bar{T}$ goes from 0 to 1,
we see that for large ${\sigma}$
\begin{equation}
\label{intlarsig}
h(\bar{T},{\sigma},{\sigma}_0) = \frac{9\pi   {\sigma}^2 g(\bar{T})} { \sqrt{3} \bar{T}^{4}} + \CO ({\sigma}^{-2})
\end{equation}
Thus the integral over ${\sigma}$ in \eqref{freeenergy3} is clearly divergent, which is expected since the system of branes is infinitely extended along the D3-brane world-volume directions. However, we can get rid of this divergence consistently as follows. First we choose to only keep fixed $\bar{T}$ and instead consider all possible values of $\Delta$. In this way we can think of the free energy $\CF( \bar{T} , \Delta(\sigma_0) ; \sigma_0 )$ as a function of $\bar{T}$ and ${\sigma}_0$. Our goal is now to compute $\CF( \bar{T} , \Delta(\sigma_0) ; \sigma_0 )$ for a large range of ${\sigma}_0$ values with ${\sigma}_0 \geq \sigma_{\rm min}(\bar{T}) = ( \bar{T}^{-8} -1)^{-1/4} $. Pick now a  ${\sigma}_0 = \sigma_{\rm cut}$ outside this range. We then consider the difference between the free energy at ${\sigma}_0$ and at $ \sigma_{\rm cut}$. This can be written as
\begin{equation}
\label{diffF}
\begin{array}{l} \ds \delta \CF ( \bar{T} , \Delta ({\sigma}_0) ; {\sigma}_0 )  \equiv  \CF(\bar{T},\Delta({\sigma}_0)  ; {\sigma}_0) - \CF(\bar{T},\Delta(\sigma_{\rm cut})  ; \sigma_{\rm cut})  \\[2mm] \ds
= \int_{{\sigma}_0}^{\sigma_{\rm cut}} d{\sigma}  h(\bar{T},{\sigma},{\sigma}_0) + \int_{\sigma_{\rm cut}}^\infty d{\sigma} \Big[ h(\bar{T},{\sigma},{\sigma}_0) - h(\bar{T},{\sigma},\sigma_{\rm cut}) \Big]
\end{array}
\end{equation}
Since the divergent part of the integrand for large ${\sigma}$ is independent of ${\sigma}_0$ the divergent part cancels out in the second integral of the RHS of Eq.~\eqref{diffF}. Moreover, since the correction to the leading part for large ${\sigma}$ in \eqref{intlarsig} goes like ${\sigma}^{-2}$ the second integral of the RHS of Eq.~\eqref{diffF} is convergent. Therefore, $\delta \CF ( \bar{T} , \Delta ({\sigma}_0) ; {\sigma}_0 )$ as defined in \eqref{diffF} is well-defined for a given $\sigma_{\rm cut}$. We can furthermore infer that the dependence on $\sigma_{\rm cut}$ only enters as an additive constant. Indeed, if we consider two values ${\sigma}_0 = a_1,a_2$ with $ a_1 < a_2 < \sigma_{\rm cut}$ we find that $\delta \CF ( \bar{T} , \Delta (a_1) ; a_1 ) - \delta \CF ( \bar{T} , \Delta (a_2) ; a_2 )$ does not depend on $\sigma_{\rm cut}$. We thus see that we can use \eqref{diffF} to measure the free energy since we only need the relative measure of the free energy between the possible branches.

\subsection{Comparison of phases}
\label{sec:comphas}

We now compare the free energy of the three distinct phases found in \cite{Grignani:2010xm} for temperatures not too large.

\begin{figure}[ht]
\centering
\includegraphics[scale=1.15]{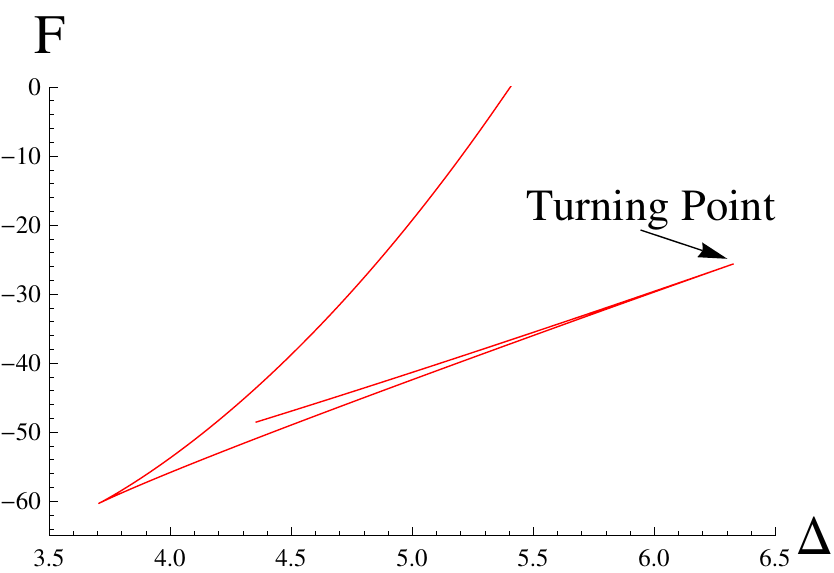}
\caption{{\small The free energy $\delta \CF$ versus $\Delta$ for $\bar{T}=0.4$ and $\kappa=1$.}
\label{fig:free1}
}
\end{figure}

We first consider the free energy for a fixed temperature. For definiteness, we consider the behavior of the free energy for the temperature $\bar{T}=0.4$. As one can see from Figure \ref{deltaT} we expect the qualitative features to be the same for the whole range of temperatures $\bar{T}$ from $0$ to $\bar{T}_b \simeq 0.8$. The $(\sigma_0,\Delta)$ diagram for $\bar{T}=0.4$ is displayed in Figure \ref{DeltaPlots}.
In Figure \ref{fig:free1} we display  $\delta \CF ( \bar{T} , \Delta( {\sigma}_0 ) ; {\sigma}_0 )$ as a $\delta \CF$ versus $\Delta$ diagram for $\bar{T} = 0.4$ (with $\sigma_{\rm cut} $ chosen so that the upper branch starts with zero free energy).
We see from these diagrams that in the range of separation distances $\Delta$ from  $\Delta_{\rm min} \simeq 3.7$ to $\Delta_{\rm max} \simeq 6.3$ the thermodynamically favored branch, $i.e.$ the branch with least free energy, is given by the branch in Figure \ref{DeltaPlots}  that goes between $\Delta_{\rm min}$ to $\Delta_{\rm max}$ with $d\Delta/d\sigma_0 < 0$. Instead when $\Delta \geq \Delta_{\rm max}$ there is only one available branch. However, as discussed below we believe this is an unstable saddlepoint. Qualitatively, this is the behavior of the phases for temperatures $0 < \bar{T} \leq \bar{T}_b \simeq 0.8$. For $\bar{T} > \bar{T}_b$ we have at most one available phase for a given $\Delta$.

Consider instead a fixed separation distance $\Delta = \Delta'$ between the two systems of branes. Increase now the temperature slowly from zero temperature. Then the thermodynamically dominant phase is the phase for which $d\Delta/d\sigma_0 < 0$, $i.e.$ the phase that goes from $\Delta_{\rm min}$ to $\Delta_{\rm max}$. Above the critical temperature $\bar{T}=\bar{T}_c$ for which $\Delta' = \Delta_{\rm max}( \bar{T})$ the phase with $d\Delta/d\sigma_0 < 0$ does not exist anymore and the only available phase is the one with $d\Delta/d\sigma_0 > 0$ that starts at $\Delta_{\rm min}$. Note that we used here that $\Delta_{\rm max} (\bar{T})$ is a monotonically decreasing function of $\bar{T}$,  see Figure \ref{deltaT}. However, we believe this phase is an unstable saddlepoint so the system should decay before reaching $\bar{T}_c$ towards another end state. This will be discussed further below in Section \ref{sec:heuristic}.

From the above considerations of the free energy we note that it has variation $d \CF = - S dT + f d\Delta$ using here that we do not allow for variations of $N$ and $k$. Consider the derivative%
\footnote{Note that since $\CF$ and $\delta \CF$ only differ by an additive constant it does not matter whether we use $\CF$ or $\delta \CF$ in the Eq.~\eqref{fdef}.
}
\begin{equation}
\label{fdef}
f = \left( \frac{\partial \CF}{\partial \Delta} \right)_{\bar{T}}
\end{equation}
$f$ is the force between the infinitely extended branes. Since it is a brane-anti-brane system the force is always positive $f > 0$. This we can in fact observe in the $\CF$ versus $\Delta$ diagram of Fig.~\ref{fig:free1}. Furthermore, since the force is continuous as we follow the curve this explains the behavior that the phases meet in cusps with zero angle.

\subsection{Energy in the extremal case}
\label{sec:extren}

\begin{figure}[ht]
\centering
\includegraphics[scale=1.15]{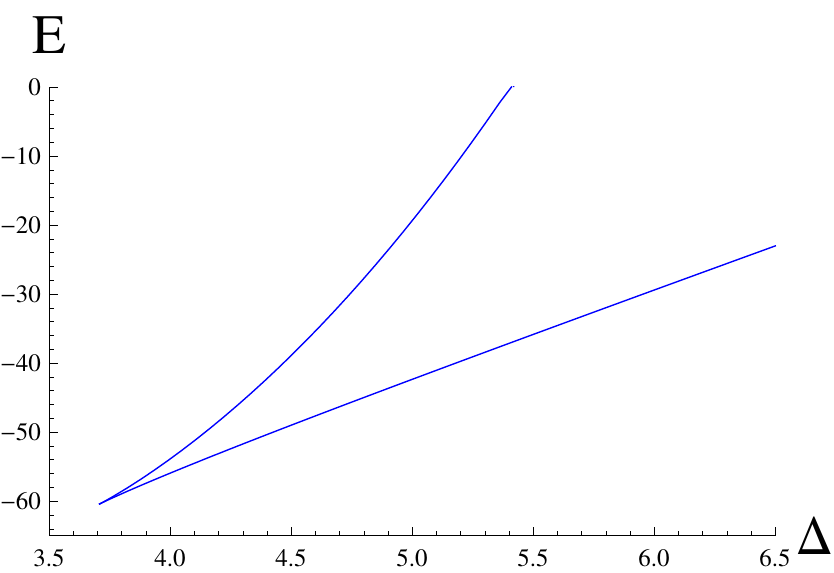}
\caption{{\small The energy $\delta E$ versus $\Delta$ for $\kappa=1$ and with $T_{\rm D3}=1$.}
\label{CMEnergy}
}
\end{figure}

It is interesting to compare the finite temperature behavior of the free energy with that of the energy in the extremal case. Also in the extremal case the energy is divergent~\cite{Grignani:2010xm} and requires a regularization. We can adopt the same regularization used above for the free energy and define a regularized energy as
\begin{equation}
\label{diffE}
\begin{array}{l} \ds \delta E (\Delta ({\sigma}_0) ; {\sigma}_0 )  \equiv  E(\Delta({\sigma}_0)  ; {\sigma}_0) - E(\Delta(\sigma_{\rm cut})  ; \sigma_{\rm cut})  \\[2mm] \ds
= \int_{{\sigma}_0}^{\sigma_{\rm cut}} d{\sigma}  h({\sigma},{\sigma}_0) + \int_{\sigma_{\rm cut}}^\infty d{\sigma} \Big[ h({\sigma},{\sigma}_0) - h({\sigma},\sigma_{\rm cut}) \Big]
\end{array}
\end{equation}
where now $h({\sigma},{\sigma}_0)$ is given by
\begin{equation}
\label{Hden1}
h({\sigma},{\sigma}_0)=8\pi T_{\rm D3}\frac{\sigma^4+\kappa^2}{\sqrt{\sigma^4-\sigma_0^4}}
\end{equation}
which corresponds to the energy density $dH/d\sigma$ in the extremal case. In this case the expression for the energy as a function of $\Delta$ can be written exactly
by eliminating $\sigma_0$ in the solution
\begin{equation}\label{X}
   z(\sigma)=\int_\sigma^{\infty}d\sigma'\frac{\sqrt{\sigma_0^4+\kappa^2}}{\sqrt{\sigma'{}^4-\sigma_0^4}}
\end{equation}
using that for fixed $\Delta$ and $\kappa$ one has
\begin{equation}\label{r02}
   \sigma_0^2=\frac{\Delta ^2\pm\sqrt{\Delta ^4-4 a^4 \kappa ^2}}{2 a^2}
   \end{equation}
where the numerical constant $a$ is given by $a^2=\frac{2\sqrt{\pi } \Gamma \left(\frac{5}{4}\right)}{\Gamma \left(\frac{3}{4}\right)}$, see \cite{Grignani:2010xm} for more details.
Since there are two solutions for $\sigma^2_0$ in \eqref{r02} these generate two branches in the energy as a function of $\Delta$, one corresponds to the ``thin throat" branch, and it is energetically favored, and the other that corresponds to the ``thick throat" branch.
This is represented in Fig.~\ref{CMEnergy}.

Comparing with the free energy diagram of Figure \ref{fig:free1} we see that the branch with highest energy/free energy behaves the same whether the temperature is turned on or not. This branch ends in both cases at $\Delta_{\rm min}$. However, the qualitative difference comes in for the remaining branches. For the zero temperature case in Figure \ref{CMEnergy} the second branch with lower energy extends all the way to infinite separation. Instead, as seen in Figure \ref{fig:free1}, for non-zero temperature there is a turning point at $\Delta_{\rm max}$. If we imagine taking the zero temperature limit of the thermal case depicted in Figure \ref{fig:free1} what happens qualitatively is that the turning point at $\Delta_{\rm max}$ is pushed towards infinity thus leaving only the two infinite branches meeting at $\Delta_{\rm min}$. This is in accordance with the fact that we get the total energy of the configuration by taking the zero temperature limit of $\CF = M - T S$ \cite{Grignani:2010xm}.

\subsection{Heuristic picture away from equilibrium}
\label{sec:heuristic}

We propose here a heuristic picture of the dynamics of the phases. In the above, we have considered static configurations in thermal and mechanical equilibrium. However, it is interesting to ask how the different phases we have found behave when allowing for dynamics.

\begin{figure}[ht]
\centering
\includegraphics[scale=0.65]{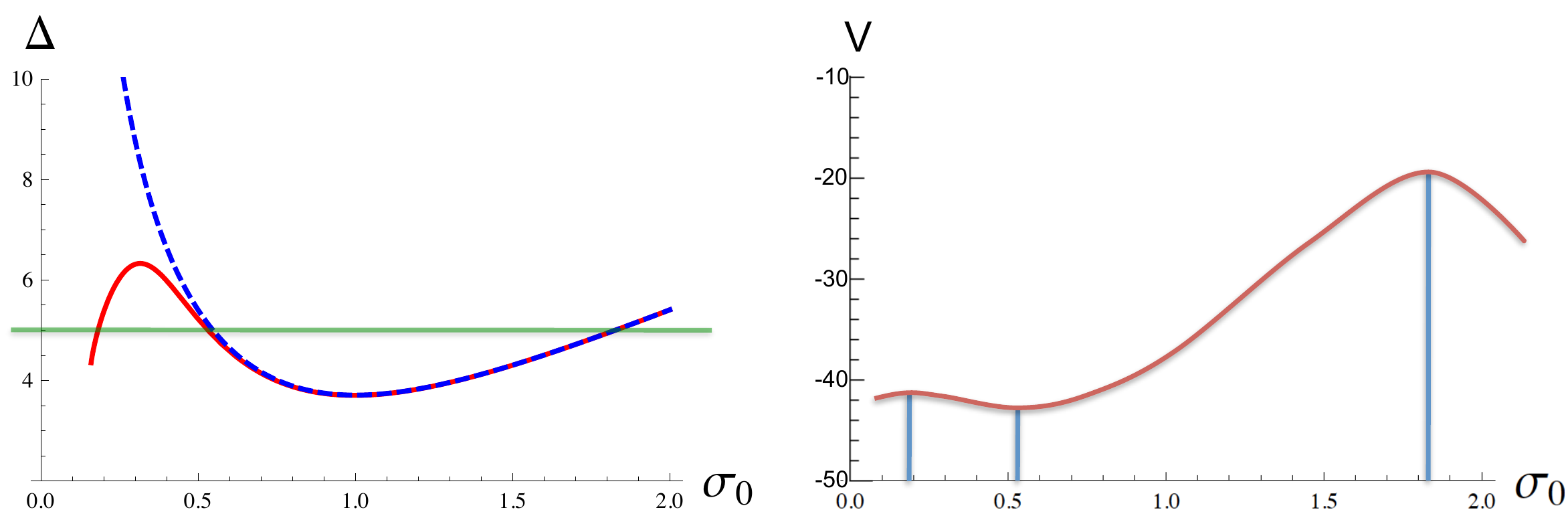}
\caption{{\small Left side: Free energy for $\bar{T}=0.4$ and $\Delta = 5$. Right side: Heuristic depiction of potential for $\bar{T}=0.4$ and $\Delta=5$.}
\label{Heuristic}
}
\end{figure}

We consider here what happens for a fixed temperature $\bar{T}=0.4$ and for different choices of values of the separation distance $\Delta$.%
\footnote{One could instead consider only one value of $\Delta$ and different choices of $\bar{T}$. However, the heuristic picture would amount to the same.} Thus, we impose the boundary conditions on the system in the form of the separation distance $\Delta$ and temperature $\bar{T}$, as measured sufficiently far away from the wormhole. The equilibrium solutions for $\bar{T}=0.4$ are displayed in Figure \ref{DeltaPlots} with free energy diagram given in Figure \ref{fig:free1}. We now fix $\Delta = 5$. This means we have three available equilibrium configurations, corresponding to three different values of the minimal radius $\sigma_0$. We illustrate this in the left part of Figure \ref{Heuristic}.

Now, one can imagine going away from equilibrium to consider configurations with different values of $\sigma_0$ but still $\Delta=5$, just as one would be able to consider a harmonic oscillator away from its equilibrium configuration. This amounts to having a potential for the system for a range of $\sigma_0$ values. We sketched such a possible potential $V$ for $\Delta=5$ and $\bar{T}=0.4$ in the right part of Figure \ref{Heuristic}. Note that the values of the potential $V$ at the equilibrium points are taken to be the values of the free energy as displayed in Figure \ref{fig:free1} since the free energy is related to the total action of the solution. We see now that in this potential we have two local maxima and one local minimum. The local minimum is, obviously, the phase with the least free energy. Thus, we expect that this phase is stable to small perturbations. Instead for the other two phases, corresponding to the two local maxima, a small perturbation could move the system increasingly further away from equilibrium.

For the large $\sigma_0$ phase, what can happen is that having a perturbation that makes $\sigma_0$ smaller would tend to take the system towards the stable equilibrium solution. So, one would presumably end up in the stable configuration. Instead, making a perturbation that would tend to increase $\sigma_0$ would result in a run away type of instability. We believe this should be in the form of a time-dependent solution where the radius of the wormhole keeps increasing and thus the brane-antibrane system will disappear. Thus, one could think of this as a brane-antibrane annihilation process. This is presumably related to open string tachyon condensation of the brane-antibrane system \cite{Sen:1998sm}.

For the small $\sigma_0$ phase, a perturbation that would increase $\sigma_0$ would presumably end up in the stable phase. Instead, a perturbation that would tend to decrease $\sigma_0$ should be such that it makes the wormhole more and more thin. We speculate that this process could end up in annihilating the F-string flux from the branes and the end point would thus be the system of infinitely extended flat branes and anti-branes, but without the wormhole.

Finally, one could contemplate what happens for other values of $\Delta$. Taking $\Delta=4$ would remove the local maximum for small $\sigma_0$ as one can see from Figure \ref{DeltaPlots}. Instead taking $\Delta=7$ both the local maximum for small $\sigma_0$ and the local minimum would disappear. Thus, one is left with an unstable phase.

\section{Thermal spike and correspondence with non-extremal string}
\label{sec:spike}

In this section we explore the question of whether there is a configuration of $k$ coincident F-strings ending on $N$ coincident D3-branes at non-zero temperature. Unlike for the zero temperature case, where this configuration for $N=1$ is described by the infinite spike BIon solution,
the analysis of the non-zero temperature case done in \cite{Grignani:2010xm} showed that the non-zero temperature analogue of the BIon solution does not allow for an infinite spike. We begin this section with some general considerations, and then we propose how the configuration in question is made by matching up a non-extremal black string solution with the throat solution found in \cite{Grignani:2010xm}. This involves finding a regime with small temperatures in which we can ignore the presence of the D3-branes in the D3-F1 system. In this regime, the D3-F1 brane bound state behaves very similar to a fundamental non-extremal string. We show this by computing the mass and entropy densities of the D3-F1 brane bound state and comparing it with the corresponding quantities for the non-extremal fundamental string.

\subsection{General considerations}
\label{sec:gencon}

The infinite spike solution of the DBI theory, first found in \cite{Callan:1997kz}, is given by
\begin{equation}
\label{infinitespike}
z(\sigma) = \frac{k T_{\rm F1}}{4\pi T_{\rm D3} \,\sigma}
\end{equation}
The interpretation of this solution is that of $k$ coincident straight F-strings ending on a single D3-brane. This brane intersection happens at a right angle, ensuring the spherical symmetry around the string.
Why is this interpretation correct? Besides that one can measure that the effective tension of the spike has the right magnitude, the other physical requirement is that sufficiently far away from the D3-brane we should not be able to see the effect of the D3-brane on the configuration anymore, and any physical measurements should be consistent with having $k$ F-strings. Naively, this could seem impossible since for any value of $z$ the radius of the throat of the spike is finite and we could thus send in a very small observer inside the throat to see that there still is a D3-brane charge present. However, we are saved by the fact that the DBI theory is not an exact theory. It is an effective theory where one integrates out the open string scale. Thus, if we are sufficiently far away, the radius of the spike is of the same magnitude as the open string scale, and we cannot see ``inside the spike" anymore. Therefore, at this point the spike solution is indistinguishable from $k$ coincident F-strings. Or, said in a different way, since the infinite spike solution \eqref{infinitespike} is not valid anymore from this point on we should match the DBI theory solution with $k$ coincident F-strings at that point. If $k$ is sufficiently large, this could involve matching up the DBI solution with a supergravity solution of $k$ coincident F-strings. Note that at the correspondence point where we match the DBI solution and the F-string supergravity solution we go beyond the validity of both solutions. However, one can extrapolate the DBI solution beyond its validity because of supersymmetry.

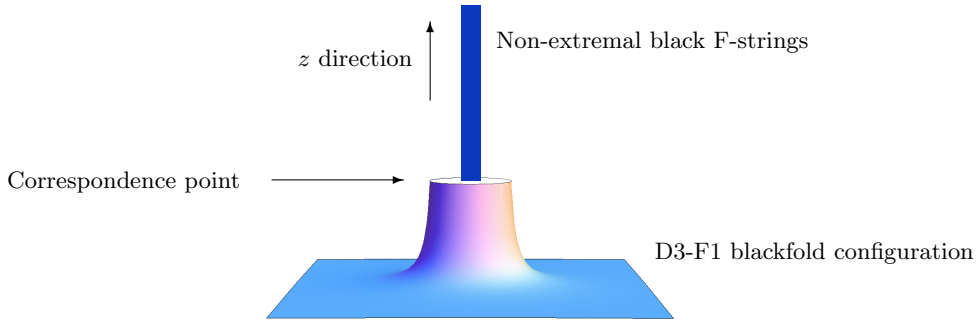
\begin{figure}[ht]
\centering
\includegraphics[scale=0.7]{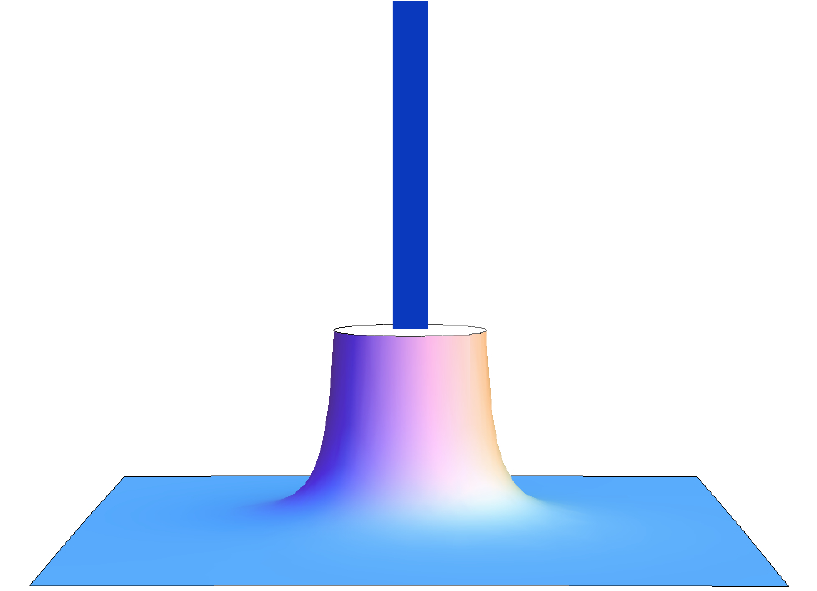}
\caption{{\small Illustration of the matching of the D3-F1 blackfold configuration and the non-extremal black F-strings at the correspondence point.}
\label{Match}
}
\begin{picture}(0,0)(0,0)
\put(-75,110){\vector(1,0){50}}
\put(-175,107){\footnotesize Correspondence point}
\put(10,160){\footnotesize Non-extremal black F-strings}
\put(70,80){\footnotesize D3-F1 blackfold configuration}
\put(-65,153){\footnotesize $z$ direction}
\put(-15,140){\vector(0,1){30}}
\end{picture}
\end{figure}

Turning now to the thermal configuration found in \cite{Grignani:2010xm}, we have that the minimal radius $\sigma_0$ is bounded from below, $\sigma_0\ge\sigma_{\rm min}$, and that $z(\sigma_0) \leq \Delta_{\rm max} / 2$ when we are not in the ``thick throat" branch that starts at $\Delta_{\rm min}$ and continues with increasing $\sigma_0$ and $\Delta$ such that $\Delta \propto \sigma_0$ for large $\sigma_0$ \cite{Grignani:2010xm}. Obviously this means that we do not have any immediate generalization of the infinite spike solution \eqref{infinitespike} in the non-zero temperature case since we do not have $z(\sigma_0)=\infty$ away from the ``thick throat" branch. However, in analogy with the zero temperature case, we shall argue below that it is possible to match the blackfold type of solution of Ref.~\cite{Grignani:2010xm}, in a certain regime, with a supergravity solution of $k$ coincident F-strings at non-zero temperature, and thereby to construct a configuration that can be interpreted as $k$ F-strings ending on $N$ D3-branes.
Just as in the zero temperature case, this involves matching up the two different solutions, the blackfold solution and the F-string supergravity solution, at a correspondence point where we have to go beyond the validity of both solutions. We argue for the matching here by using the blackfold solution beyond its validity to show that its physical behavior is very close to that of $k$ non-extremal F-strings, and the slight numerical differences can be explained by the fact that we do not have supersymmetry to protect us.
We illustrate the matching of the blackfold solution and non-extremal black F-string solution at the correspondence point in Figure \ref{Match}.

\subsection{Correspondence point for matching of throat to F-strings}

In \cite{Grignani:2010xm} we found a solution using the blackfold approach for a spherically symmetric configuration of $N$ D3-branes with a throat supported by an F-string flux such that the F-string charge measured over each spherical surface is $k$. The throat ends in a minimal two-sphere of radius $\sigma_0$. In this section we find a regime of the solution of Ref.~\cite{Grignani:2010xm}  in which it behaves like $k$ non-extremal F-strings at the end of the throat, $i.e.$ at $\sigma=\sigma_0$.%
\footnote{In order to compare our findings with the results for a non extremal fundamental string we work in this section in terms of the quantities $(T,N,k)$, $T_{\rm F1}$ and $T_{\rm D3}$ instead of $\kappa$ and $\bar T$.} This enables us to match the D3-F1 blackfold configuration to $k$ non-extremal black F-strings. As we explain below, this means that the end of the throat $\sigma=\sigma_0$ serves as the correspondence point between the two different solution, as illustrated in Figure \ref{Match}.

Since we are aiming to generalize the extremal infinite spike solution of  \cite{Callan:1997kz} to non-zero temperature, as discussed above in Section \ref{sec:gencon}, we approach this problem in the low temperature limit in which we are close to extremality. We begin therefore this section by examining the thermodynamics of the supergravity solution of $k$ non-extremal F-strings at low temperature, and subsequently find a corresponding regime of the non-extremal D3-F1 solution of  \cite{Grignani:2010xm}.

\subsubsection*{Non-extremal black F-strings at low temperature}

The supergravity solution for $k$ coincident non-extremal black F-strings lying along the $z$ direction is
\begin{equation}
\begin{array}{c} \ds
ds^2 = H^{-1} ( - f dt^2 + dz^2 ) + f^{-1} dr^2 + r^2 d\Omega_7^2
\\[2mm] \ds
e^{2\phi} = H^{-1} \spa B_{0z} = H^{-1} - 1 \spa H = 1+  \frac{r_0^6 \sinh^2 \bar{\alpha}}{r^6} \spa f = 1 - \frac{r_0^6}{r^6}
\end{array}
\end{equation}
here written in the string frame. This supergravity solution is a good description of $k$ non-extremal F-strings for $k\gg1$ and $g_s^2 k \gg 1$.%
\footnote{That $g_s^2 k \gg 1$ follows from demanding that the curvature length scale $(G T_{\rm F1} k)^{1/6} \sim ( g_s^2 k )^{1/6} l_s $ for the supergravity solution of $k$ coincident extremal F-strings is larger than the string scale.}
From this the mass density along the $z$ direction can be found using Ref.~\cite{Harmark:2004ch} %
\footnote{We write Eq.~\eqref{MF1} in terms of $T_{\rm D3}$ rather than the 10D Newtons constant $G$. This is done using the relation $16 \pi G=(2\pi)^7 l_s^8 g_s^2=(2\pi)/T_{\rm D3}^2$.}
\begin{equation}\label{MF1}
\frac{dM_{\rm F1}}{dz}= \frac{3^5 T_{\rm D3}^2 (1+6\cosh^2\bar{\alpha})}{2^7 \pi^3 T^6 \cosh^6 \bar{\alpha}} \spa k^2 = \frac{3^{12} T_{\rm D3}^4 ( \cosh^2 \bar{\alpha} - 1)}{2^{12} \pi^6 T_{\rm F1}^2 T^{12} \cosh^{10} \bar{\alpha}} \, ,
\end{equation}
where we eliminated $r_0$ in favor of the temperature $T$ and the second equation follows from charge quantization. Combining the two equations one can eliminate $\bar{\alpha}$ and write the mass density in terms of $k$, $T$, $T_{\rm D3}$ and $T_{\rm F1}$. Note that in solving the second equation of \eqref{MF1} for $\bar{\alpha}$ we choose the branch connected to the extremal solution. For small temperatures one can expand the mass density as follows
\begin{equation}\label{MF1exp}
\frac{dM_{\rm F1}}{dz} = T_{\rm F1} k+\frac{16 \left(T_{\rm F1} k \,\pi \right)^{3/2} {T}^3}{81
   T_{\rm D3}}+\frac{40 T_{\rm F1}^2 k^2 \pi ^3 {T}^6}{729 T_{\rm D3}^2}+\mathcal{O}\left(T^9\right)
\end{equation}
This is the mass density of $k$ coincident F-strings for low temperatures so that the F-strings are close to extremality as found in the regime $k\gg 1$. We see the leading part as expected is given by $k$ times the extremal string tension $T_{\rm F1}$. We also read from this expression that being near extremality means that we are in the regime
\begin{equation}
\frac{\sqrt{T_{\rm F1} k}\, T^3}{T_{\rm D3}}\ll 1
\label{stringpert}
\end{equation}
In terms of the illustration in Figure \ref{Match} we can say that \eqref{MF1exp} is the mass density in the $z$ direction for the non-extremal black F-strings in the upper part of the drawing, away from the correspondence point.

\subsubsection*{Finding matching regime in the thermal D3-F1 configuration}

Having found the low temperature behavior \eqref{MF1exp} of the mass density of $k$ F-strings at low temperature near extremality we now look for a regime of the thermal D3-F1 blackfold configuration of  \cite{Grignani:2010xm} in which its mass density is similar at the correspondence point at the end of the throat $\sigma=\sigma_0$. With such a regime in hand, we can match the D3-F1 blackfold configuration and the non-extremal F-string supergravity solution, as explained above in Section \ref{sec:gencon}.

Using \eqref{Msolution} along with the D3-F1 blackfold configuration given by \eqref{thesolution}-\eqref{T_kappa_delta} we can compute the mass density%
\footnote{Note that we have a factor two less in the mass since here we do not include the mirror of the solution.}
\begin{equation}\label{dMdz}
       \frac{dM}{dz}=\frac{2 T_{\rm D3}^2}{\pi T^4}\,
        \frac{F(\sigma)}{F(\sigma_0)}\,\sigma^2\frac{4 \cosh^2 \alpha+ 1}{\cosh^4 \alpha}\, .
\end{equation}
where we divided by the derivative of the
solution, $z'(\sigma)$, to construct the mass density with respect to the transverse direction $z(\sigma)$ to the D3-brane. We now expand this in the regime $\bar{T} \ll 1$ both since we want to match the result to the small temperature expansion of the non-extremal F-strings \eqref{MF1exp} but also since $\bar{T}=1$ is the maximal temperature for the D3-brane without F-string flux and thus a regime in which the D3-branes are suppressed with respect to the F-strings would have to be far from $\bar{T} =1$. Using the expansion for small temperature of the solution found in \cite{Grignani:2010xm} we obtain a perturbative expansion of the mass density \eqref{dMdz} in powers of the temperature $T$ for $\bar{T}\ll 1$
\begin{equation}
 \label{dMdzexp}
\left.\frac{dM}{dz}\right|_{\sigma=\sigma_0}= T_{\rm F1} k \sqrt{1+x^2} + \frac{3 \pi T_{\rm F1}^2 k^2}{32 T_{\rm D3}^2 \sigma_0^2} (1+x^2) T^4 + \frac{7\pi^2 T_{\rm F1}^3 k^3 }{512 T_{\rm D3}^4 \sigma_0^4 } (1+x^2)^{\frac{3}{2}} T^8 + \CO ( T^{12} ) \, ,
\end{equation}
where we defined the quantity
\begin{equation}
\label{xdef}
x \equiv \frac{4\pi T_{\rm D3} \sigma_0^2 N}{T_{\rm F1} k} \, .
\end{equation}
Note that from \eqref{dMdzexp} it is apparent that we need a stronger requirement than $\bar{T} \ll 1$ for the correction terms in the expansion \eqref{dMdzexp} to be small, namely that
\begin{equation}
\label{expansioncond}
\bar{T}^4 \ll \frac{x}{\sqrt{1+x^2}} \, .
\end{equation}
We now demand that we are in a regime in which the mass density \eqref{dMdzexp} is dominated by the presence of the $k$ F-strings to (and including) order $T^8$. This basically means that any term with $N$ dependence should be suppressed in comparison with the terms with only $k$ dependence. For the leading order term in \eqref{dMdzexp} this clearly means that $x \ll 1$. Since we recognize $T_{\rm F1} k \sqrt{1+x^2}$ as the $1/2$ BPS mass density formula for a D3-F1 brane bound state we see that $x \ll 1$ physically is the requirement that the F-strings dominate in the extremal bound state. Turning to the next order in the temperature, the $T^4$ term in \eqref{dMdzexp}, we see that the requirement of the dominance of the F-strings means $T_{\rm F1} k \, x^2    \ll T_{\rm F1}^2 k^2 T^4 /( T_{\rm D3}^2 \sigma_0^2 )$. We can write this condition as $x^3 \ll \bar{T}^4$. We see that this condition implies the $T^0$ condition $x\ll 1$ since our starting point is that $\bar{T} \ll 1$. Going on to the order $T^8$ we should require that $T^8 T_{\rm F1}^3 k^3 / ( T_{\rm D3}^4 \sigma_0^4 )$ dominates over both $x^2 T_{\rm F1} k$ and $x^2 T^4 T_{\rm F1} k^2 / ( T_{\rm D3}^2 \sigma_0^2 )$. While the latter reduces to the condition $x^3 \ll \bar{T}^4$ the first one gives instead the condition $x^2 \ll \bar{T}^4$
and we see that this condition implies both the $T^0$ order condition $x\ll 1$ and the $T^4$ order condition $x^3 \ll \bar{T}^4$.
If we now consider the condition \eqref{expansioncond} on the expansion of \eqref{dMdzexp} we note that for $x \ll 1$ this becomes $\bar{T}^4 \ll x$ which is consistent with the $x^2 \ll \bar{T}^4$ condition.
Thus, in summary, demanding that the physics of the $k$ F-strings should dominate over the $N$ D3-branes at the end of the throat $\sigma=\sigma_0$ to (and including) order $T^8$ means that we should be in a regime with
\begin{equation}
\label{xTregime}
x^2 \ll \bar{T}^4 \ll x\, .
\end{equation}
If one should include higher order corrections in the temperature expansion one would get stronger requirements on $x$. Assuming \eqref{xTregime} the mass density \eqref{dMdzexp} now becomes
\begin{equation}
 \label{dMdzexp2}
\left.\frac{dM}{dz}\right|_{\sigma=\sigma_0}= T_{\rm F1} k + \frac{3 \pi T_{\rm F1}^2 k^2}{32 T_{\rm D3}^2 \sigma_0^2} T^4 + \frac{7\pi^2 T_{\rm F1}^3 k^3 }{512 T_{\rm D3}^4 \sigma_0^4 } T^8 + \CO ( T^{12} ) \, ,
\end{equation}
Comparing this to the mass density for the F-strings \eqref{MF1exp}
we see that in order for the thermal D3-F1 configuration at $\sigma=\sigma_0$ to behave like $k$ F-strings near extremality we should take $\sigma_0$ to have the following dependence on the temperature
\begin{equation}
\sigma_0=\left(\frac{\sqrt{T_{\rm F1} k}}{T_{\rm D3}}\right)^{\frac{1}{2}}\, \sqrt{T}\left(s_0+s_1\, \frac{\sqrt{T_{\rm F1} k}}{T_{\rm D3}}\,T^{3}+\mathcal{O}(T^6)\right)
\label{sigmaTdep}
\end{equation}
where $s_0$ and $s_1$ are numerical coefficients which can be determined by requiring that the $T^3$ and $T^6$ terms in Eqs.~\eqref{MF1exp} and \eqref{dMdzexp2} agree. We find the precise coefficients below. Note that from the regime \eqref{xTregime} we have $\bar{T}^4 / x \ll 1$ which becomes the condition \eqref{stringpert} when we insert \eqref{sigmaTdep}.

The condition \eqref{xTregime} can be written equivalently as $\bar{T}^4 \ll x \ll \bar{T}^2$. Since from \eqref{xdef} we see that $x \sim \sigma_0^2 / \kappa$ we can also write it as $\bar{T}^2 \ll \sigma_0/\sqrt{\kappa} \ll \bar{T}$. This shows that we are well above $\sigma_{\rm min}$ but also below $\sigma_0 / \sqrt{\kappa} \sim \bar{T}^{2/3}$ for which $\Delta \sim \Delta_{\rm max}$. If we insert \eqref{sigmaTdep} in $\sigma_0/\sqrt{\kappa} \ll \bar{T}$ this condition can be written as $N T_{\rm D3} / ( k T_{\rm F1} ) \ll T^2$. This is consistent with \eqref{stringpert} when $N^3/k^2 \ll g_s$.

\subsubsection*{Validity of the probe approximation and the correspondence point}

Having found the regime \eqref{xTregime} with $\sigma_0$ given by \eqref{sigmaTdep} we have identified a regime in which the D3-F1 blackfold configuration of Ref.~\cite{Grignani:2010xm} at the end of the throat $\sigma=\sigma_0$ behaves as $k$ non-extremal F-strings near extremality. As illustrated in Figure \ref{Match}, the end of the throat $\sigma=\sigma_0$ is what we call the correspondence point. In our case we take this to signify that it is a point in which it is possible to match the two descriptions although the validity of the two descriptions does not overlap. One should therefore extrapolate the two descriptions in order to match them. Above we did this by finding the regime in which the D3-F1 blackfold description of the system looks like the non-extremal F-string description at the correspondence point.

That the validity of the approximation of the D3-F1 blackfold configuration breaks down when we are in the regime \eqref{xTregime}, \eqref{sigmaTdep} is seen as follows. In \cite{Grignani:2010xm} it was found that the probe approximation is valid provided $r_c \ll \sigma_0$ where $r_c$ is the charge radius of the D3-F1 brane bound state (see Section 4.4 of \cite{Grignani:2010xm}). Note that from \eqref{xTregime} we have that $x\ll 1$ which means that $\sigma_0 \ll \sqrt{\kappa}$. Using this one finds that the condition $r_c \ll \sigma_0$ reduces to $k T_{\rm F1} \ll \sigma_0^6 T_{\rm D3}^2$. Inserting \eqref{sigmaTdep} we then find that the blackfold approximation requires
\begin{equation}
\label{BFapprox}
\frac{\sqrt{k T_{\rm F1}}}{T_{\rm D3}} T^3 \gg 1
\end{equation}
which is seen to be exactly the opposite of the requirement \eqref{stringpert}. Thus, clearly we are beyond the validity of the blackfold description. However, this is in fact consistent with what we would like to do. If we for a moment imagine that we could reach a regime in which both the D3-F1 blackfold description and the non-extremal F-string description are valid, we would run in to problems. Indeed, an observer smaller than the size of the throat at $\sigma=\sigma_0$ would be able to see that there is a hole in the middle of the string and that there still is a D3-brane charge present. Instead, that the blackfold regime breaks down means that the backreaction to the system becomes strong before we reach $\sigma=\sigma_0$ and this can thus close off the hole in the middle of the throat and thereby also cancel out the D3-brane charge.

Of course, the fact that we go far beyond the validity of the blackfold description also means that it is limited what we can conclude from our approach. However, at the same time the breakdown of the blackfold description happening precisely in the regime in which we get a behavior similar to the F-strings is perhaps the clearest indication that our match between the D3-F1 blackfold and the non-extremal F-strings is possible.

To find the exact configuration in which $k$ non-extremal F-strings dissolve into $N$ non-extremal D3-branes would be a very difficult problem. One could try to add order by order corrections on the blackfold side, taking into account the backreaction, and at the same time add corrections to the F-string side, although it is not clear at present how one should do that. Note here that even in the extremal case such a configuration has not been found, thus as a first approach one should start with that.

\subsubsection*{Extrapolating the D3-F1 configuration}

Finally, we end this section with seeing how far we can take the extrapolation of the D3-F1 blackfold configuration in making it look like $k$ non-extremal F-strings. Note that a more accurate calculation should take into account that one should extrapolate both descriptions instead of only the D3-F1 blackfold description. However, we do not know how the F-string side changes as we get closer to the D3-branes.

We already remarked above that the mass density \eqref{dMdzexp2} can be matched to \eqref{MF1exp} with the precise coefficients up to (and including) order $T^6$ by fixing $s_0$ and $s_1$ in \eqref{sigmaTdep}. We find the numbers $s_0 = {9 \sqrt{3} } / (16 \sqrt{2} \sqrt[4]{\pi })$ and $s_1 = {43 \pi ^{5/4} }/(1728 \sqrt{6})$. This can be continued to higher orders as well. However, fixing $dM/dz$ as function of $k$ and $T$ does not fully characterize the thermodynamics of the string. This requires matching another quantity as well. We consider here the entropy density $dS/dz$. For $k$ non-extremal F-strings we find
\begin{equation}
    \frac{dS_{\rm F1}}{dz} = \frac{8 \left(T_{\rm F1} k \pi \right)^{3/2} {T}^2}{27
       T_{\rm D3}}+\frac{16 T_{\rm F1}^2 k^2 \pi ^3 {T}^5}{243 T_{\rm D3}^2}+\mathcal{O}\left(T^8\right) \, , \label{F1entropy}
\end{equation}
in the regime \eqref{stringpert}. One can now play the same game as for $dM/dz$ and fix $s_0$ and $s_1$ in \eqref{sigmaTdep} so that the D3-F1 blackfold configuration at $\sigma=\sigma_0$ reproduces $dS_{\rm F1}/dz$. As one could expect, the values of $s_0$ and $s_1$ that one needs to match $dM/dz$ and $dS/dz$ are different. Thus, one cannot simultaneously match the two quantities exactly. However, this would also be highly surprising since we are beyond the validity of the blackfold approximation, as discussed above. Nevertheless, it is interesting to compare how different the first correction to the extremal configuration is since for the extremal configuration we have supersymmetry and all the quantities thus match exactly. Computing $s_0$ for the two quantities we find
\begin{equation}
s_0 |_{\frac{dM}{dz}} \simeq 0.69 \spa  s_0 |_{\frac{dS}{dz}} \simeq 0.65 \, .
\end{equation}
We see that they are about $6\%$ different. This is encouraging, as it basically means that we could reproduce the F-string thermodynamics within $6\%$ accuracy which is a rather good accuracy given that we are extrapolating the blackfold description beyond its validity.

\section*{Acknowledgments}

We thank Jan de Boer, Nadav Drukker, Vasilis Niarchos, Gordon Semenoff, Larus Thorlacius and especially Roberto Emparan
for useful discussions. TH thanks NBI for hospitality and MO and NO are grateful to Nordita for hospitality
during the workshop on ``Integrability in String and Gauge Theories; AdS/CFT Duality and its Applications''.


\addcontentsline{toc}{section}{References}

\providecommand{\href}[2]{#2}\begingroup\raggedright\endgroup


\begin{thebibliography}{1}

\bibitem{Grignani:2010xm}
G.~Grignani, T.~Harmark, A.~Marini, N.~A. Obers, and M.~Orselli, ``{Heating up
  the BIon},''
\href{http://arxiv.org/abs/1012.1494}{{\tt arXiv:1012.1494 [hep-th]}}.


\bibitem{Emparan:2007wm}
R.~Emparan, T.~Harmark, V.~Niarchos, N.~A. Obers, and M.~J. Rodriguez, ``{The
  Phase Structure of Higher-Dimensional Black Rings and Black Holes},''
  \href{http://dx.doi.org/10.1088/1126-6708/2007/10/110}{{\em JHEP} {\bf 10}
  (2007)  110},
\href{http://arxiv.org/abs/0708.2181}{{\tt arXiv:0708.2181 [hep-th]}}.
%
R.~Emparan, T.~Harmark, V.~Niarchos, and N.~A. Obers, ``{World-Volume Effective
  Theory for Higher-Dimensional Black Holes},''
  \href{http://dx.doi.org/10.1103/PhysRevLett.102.191301}{{\em Phys. Rev.
  Lett.} {\bf 102} (2009)  191301},
\href{http://arxiv.org/abs/0902.0427}{{\tt arXiv:0902.0427 [hep-th]}}.
%
R.~Emparan, T.~Harmark, V.~Niarchos, and N.~A. Obers, ``{Essentials of
  Blackfold Dynamics},'' \href{http://dx.doi.org/10.1007/JHEP03(2010)063}{{\em
  JHEP} {\bf 03} (2010)  063},
\href{http://arxiv.org/abs/0910.1601}{{\tt arXiv:0910.1601 [hep-th]}}.
%
R.~Emparan, T.~Harmark, V.~Niarchos, and N.~A. Obers, ``{New Horizons for Black
  Holes and Branes},'' \href{http://dx.doi.org/10.1007/JHEP04(2010)046}{{\em
  JHEP} {\bf 04} (2010)  046},
\href{http://arxiv.org/abs/0912.2352}{{\tt arXiv:0912.2352 [hep-th]}}.
%
R.~Emparan, T.~Harmark, V.~Niarchos, and N.~A. Obers, ``{Blackfold approach for
  higher-dimensional black holes},''
{\em Acta Phys. Polon.} {\bf B40} (2009)  3459--3506.
R.~Emparan, T.~Harmark, V.~Niarchos, and N.~A. Obers, ``{Charged Blackfolds},''
  {\em To appear}  .


\bibitem{Callan:1997kz}
C.~G. Callan and J.~M. Maldacena, ``{Brane dynamics from the Born-Infeld
  action},'' \href{http://dx.doi.org/10.1016/S0550-3213(97)00700-1}{{\em Nucl.
  Phys.} {\bf B513} (1998)  198--212},
\href{http://arxiv.org/abs/hep-th/9708147}{{\tt arXiv:hep-th/9708147}}.

\bibitem{Gibbons:1997xz}
G.~W. Gibbons, ``{Born-{Infeld} particles and Dirichlet {$p$}-branes},''
  \href{http://dx.doi.org/10.1016/S0550-3213(97)00795-5}{{\em Nucl. Phys.} {\bf
  B514} (1998)  603--639},
\href{http://arxiv.org/abs/hep-th/9709027}{{\tt arXiv:hep-th/9709027}}.
%
L.~Thorlacius, ``{Born-Infeld string as a boundary conformal field theory},''
  \href{http://dx.doi.org/10.1103/PhysRevLett.80.1588}{{\em Phys. Rev. Lett.}
  {\bf 80} (1998)  1588--1590},
\href{http://arxiv.org/abs/hep-th/9710181}{{\tt arXiv:hep-th/9710181}}.
%
R.~Emparan, ``{Born-Infeld strings tunneling to {D}-branes},''
  \href{http://dx.doi.org/10.1016/S0370-2693(98)00107-5}{{\em Phys. Lett.} {\bf
  B423} (1998)  71--78},
\href{http://arxiv.org/abs/hep-th/9711106}{{\tt arXiv:hep-th/9711106}}.

\bibitem{Hartnoll:2006hr}
S.~A. Hartnoll and S.~Prem~Kumar, ``{Multiply wound Polyakov loops at strong
  coupling},'' \href{http://dx.doi.org/10.1103/PhysRevD.74.026001}{{\em Phys.
  Rev.} {\bf D74} (2006)  026001},
\href{http://arxiv.org/abs/hep-th/0603190}{{\tt arXiv:hep-th/0603190}}.
%
G.~Grignani, J.~L. Karczmarek, and G.~W. Semenoff, ``{Hot Giant Loop
  Holography},'' \href{http://dx.doi.org/10.1103/PhysRevD.82.027901}{{\em Phys.
  Rev.} {\bf D82} (2010)  027901},
\href{http://arxiv.org/abs/0904.3750}{{\tt arXiv:0904.3750 [hep-th]}}.


\bibitem{Sen:1998sm}
A.~Sen, ``{Tachyon condensation on the brane antibrane system},'' {\em JHEP}
  {\bf 08} (1998)  012,
\href{http://arxiv.org/abs/hep-th/9805170}{{\tt arXiv:hep-th/9805170}}.

\bibitem{Harmark:2004ch}
T.~Harmark and N.~A. Obers, ``{General definition of gravitational tension},''
  \href{http://dx.doi.org/10.1088/1126-6708/2004/05/043}{{\em JHEP} {\bf 05}
  (2004)  043},
\href{http://arxiv.org/abs/hep-th/0403103}{{\tt arXiv:hep-th/0403103}}.

\end{thebibliography}
\end{document}